\documentclass[twocolumn,aps,showpacs,preprintnumbers,superscriptaddress,amsmath,amssymb,prl]{revtex4-1}  
\usepackage{graphicx}  
\usepackage{dcolumn}   
\usepackage{bm}        
\usepackage{color}

\def\natp{Nature\ Photon. }
\def\ptl{IEEE\ Phot.\ Technol.\ Lett. }
\def\jlt{J.\ Lightwave\ Technol. }
\def\pla{Phys.\ Lett.\ A }
\def\nl{Nano\ Lett. }
\def\pccp{Phys.\ Chem.\ Chem.\ Phys. }
\def\prep{Phys.\ Rep. }
\def\opex{Opt.\ Express }
\def\nan{Nanotechnol. }
\def\jap{J.\ Appl.\ Phys. }

\begin{document}

\title{Nonlinear Surface-Plasmon Whispering-Gallery Modes in Metallic Nanowire Cavities}
\date{\today}

\author{C.~G.~Biris}
\affiliation{Department of Electronic and Electrical Engineering, University College London,
Torrington Place, London WC1E 7JE, United Kingdom}
\altaffiliation{C. G. Biris is now with the High Performance Computing Centre, West University of Timisoara, Bd. V. Parvan,
Nr. 4, 300233 Timisoara, Romania.}

\author{N.~C.~Panoiu}
\affiliation{Department of Electronic and Electrical Engineering, University College London,
Torrington Place, London WC1E 7JE, United Kingdom} \affiliation{Thomas Young Centre, London Centre
for Nanotechnology, University College London, 17-19 Gordon Street, London, WC1H 0AH, UK}

\begin{abstract}
We demonstrate that the surface second-harmonic generation can lead to the formation of nonlinear
plasmonic whispering-gallery modes (WGMs) in microcavities made of metallic nanowires. Since these
WGMs are excited by induced surface nonlinear dipoles, they can be generated even when they are
not coupled to the radiation continuum. Consequently, the quality factor of these nonlinear modes
can be as large as the theoretical limit imposed by the optical losses in the metal. Remarkably,
our theoretical analysis shows that nonlinear plasmonic WGMs are characterized by fractional
azimuthal modal numbers. This suggests that the plasmonic cavities investigated here can be used
to generate multi-color optical fields with fractional angular momentum. Applications to plasmonic
sensors are also discussed.
\end{abstract}

\maketitle

Whispering-gallery (WG) optical microresonators play a key role in many photonic applications,
including microcavity lasers \cite{mls92apl,ctn11np}, optical filters and modulators
\cite{dcd02ptl,rsz02jlt,km07ol}, lasers \cite{kdh92ol,sth96pra,pmk04apl}, mechanical, chemical,
and biological sensors \cite{msi04oc,ssr04pccp,vbl02apl}, nonlinear optics
\cite{bgi89pla,ksv04prl}, and quantum optics \cite{cel91prl,vfg98pra,aww01nl} (for a review of the
properties of WG microresonators and their applications see Refs. \cite{mi06stq,ia06stq}). These
microresonators are essential elements in such a broad array of applications due to the unique
properties of their optical modes, namely the strong confinement and enhancement of the optical
field, and the rich spectrum of available choices for optical materials and configurations that
can be used to fabricate WG microresonators. In particular, closely spaced WG modes (WGMs) with
ultra-high $Q$-factor and small modal volume can be easily achieved, $e.g.$, by using dielectric
microspheres, capillary cavities, photonic crystal cavities, and liquid droplets. However, since
WGMs are formed by total internal reflection at closed, curved interfaces, reducing the modal
volume of WGMs of dielectric based microresonators beyond a few cubic wavelengths is a major
challenge.

The size of WGMs can be significantly decreased by using plasmonic resonators, such as metallic
nanoparticles and cavities \cite{csb06prl,mos09nat,nzb09nat,nsb10natp,vgp09nl,xzl10prl}, as they
can confine the optical field to a domain that is comparable or significantly smaller than the
optical wavelength \cite{bde03n,zsm05pr,vme10rmp,m07book}. While $Q$-factors of plasmonic
resonators are smaller than those of dielectric cavities, chiefly due to the optical losses in the
metal, plasmonic cavities with $Q$-factors larger than $10^{3}$ have been recently demonstrated
\cite{mos09nat,nsb10natp,bp1011oenan}. As a result, it is possible to achieve plasmonic resonators
with ratio between the $Q$-factor and the modal volume, which is a key figure of merit of the
performance of optical cavities, as large as that of dielectric cavities. However, unlike
dielectric cavities, plasmonic resonators can enhance the optical near-field at the metal surface
by more than two orders of magnitude, which, in conjunction with the strong sensitivity of the
near-field to changes of the physical and chemical properties of metal surfaces, can be
exploited in many applications pertaining to light control at deep subwavelength scale. This is
particularly important in the case of nonlinear optics applications, as then the local field
enhancement effect is greatly amplified by the nonlinear character of the optical interaction.

\begin{figure}[t]
\includegraphics*[width=7cm]{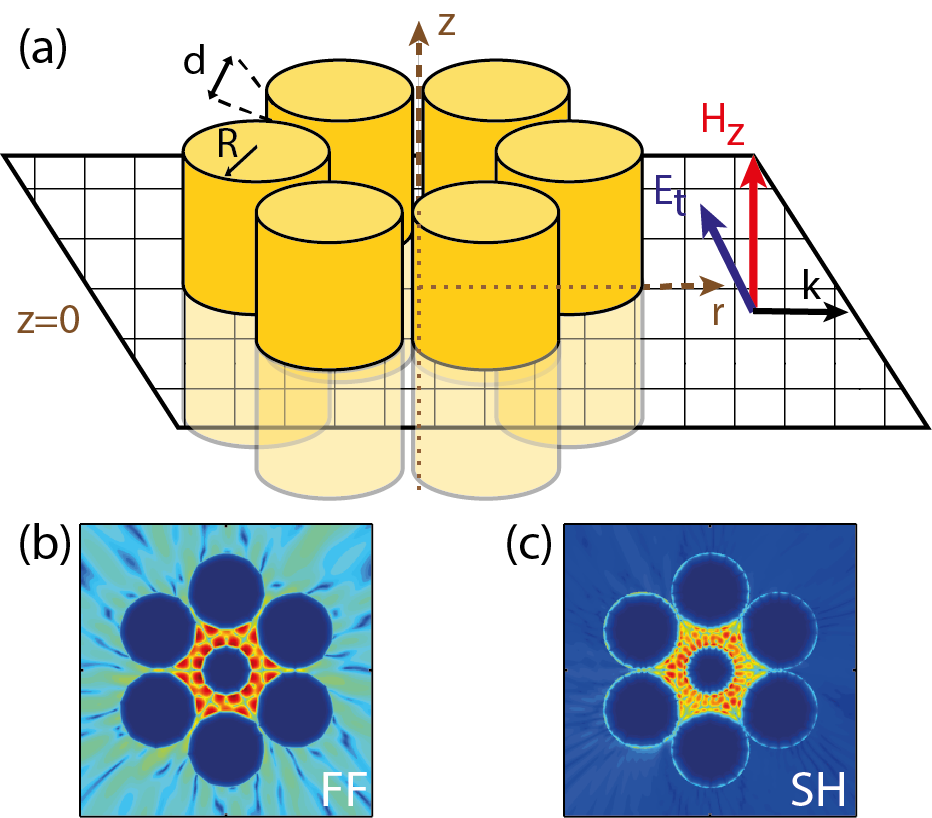}
\caption{\label{fig:geometry} (a) Schematics of the plasmonic cavity and field polarization. (b),
(c) Amplitude of the magnetic field of a WGM of a cavity with $R=1500~\mathrm{nm}$,
$d=10~\mathrm{nm}$, and internal cylindrical inclusion with $R_{i}=1000~\mathrm{nm}$. The excitation
is a multipole source with $m_0=2$ and $\lambda_{\mathrm{FF}}=989~\mathrm{nm}$.}
\end{figure}

In this Letter we reveal for the first time the existence of nonlinear plasmonic WGMs formed as
the result of second-harmonic generation (SHG) in plasmonic cavities made of metallic nanowires.
Importantly, these nonlinear WGMs do not couple with the free-space radiative modes, which allows
one to achieve $Q$-factors as large as the theoretical limit imposed by the optical loss in the
metal. A typical geometry of the cavities considered in our study is shown in Fig.
\ref{fig:geometry}. The cavity consists of $N$ infinite, parallel nanowires made of Au, which are
embedded in a dielectric medium with permittivity $\epsilon_{b}$. The nanowires have radius $R$
and are separated by a distance, $d$. In a more general case, we assume that at the center of the
cavity there is another nanowire with radius $R_{i}$.

\textit{Theoretical analysis and numerical method}.---To study the nonlinear WGMs of plasmonic
cavities we use a well known numerical algorithm based on the multiple-scattering matrix (MSM)
method \cite{ftm94josaa}, modified to account for the surface SHG \cite{bp10prb}. To begin with,
we assume that the plasmonic cavity is excited by a TE polarized field. For the TM polarization
the electric field is parallel to the surface of the nanowires and thus no SH is generated. Thus,
we write the excitation field (a plane-wave in our case), $\vert \psi_{\mathrm{ex}}\rangle$, at
the FF, $\omega$, as a Fourier-Bessel series,
\begin{equation}
  \label{eq:incoming}
  \vert \psi_{\mathrm{ex}}\rangle = \sum_{m=-\infty}^{\infty} a_{m}^{(0)} \vert Jm\rangle,
\end{equation}
where $a_{m}^{(0)}$ are the expansion coefficients of the excitation field. For simplicity, we
have used Dirac notation, with the excitation field in the real space being
$\langle\mathbf{r}\vert \psi_{\mathrm{ex}}\rangle=H_{\mathrm{ex},z}(r,\phi)$, the
\textit{z}-component of the magnetic field of the incident wave. In this notation,
$\langle\mathbf{r}\vert Jm\rangle=J_{m}(kr) e^{im\phi}$ are multipole functions of \textit{m}-th
order ($k=\omega/c$), with $J_{m}(x)$ being cylindrical Bessel functions of first kind. For a
plane wave excitation, $\langle\mathbf{r}\vert
\psi_{\mathrm{ex}}\rangle=H_{0}e^{-i\mathbf{k}\cdot\mathbf{r}}$, the coefficients
$a_{m}^{(0)}=(-i)^{m}H_{0}$, so that the symmetry relation $a_{m}^{(0)}=(-1)^{m}a_{-m}^{(0)}$
holds. The total field can be written as the sum between the excitation and scattered fields,
$\vert \psi_{\mathrm{tot}}\rangle=\vert \psi_{\mathrm{ex}}\rangle+\vert
\psi_{\mathrm{sc}}\rangle$. The scattered field is expanded as
\begin{equation}
  \label{eq:scattered}
  \vert \psi_{\mathrm{sc}}\rangle = \sum_{\alpha=1}^{N}\sum_{m=-\infty}^{\infty} b_{m}^{(\alpha)} \vert
  Hm\rangle_{\alpha}.
\end{equation}
Here, $b_{m}^{(\alpha)}$ are the expansion coefficients of the scattered field and
$\langle\mathbf{r}\vert Hm\rangle_{\alpha}=H_{m}^{(2)}(kr_{\alpha}) e^{im\phi_{\alpha}}$ are
multipole functions of \textit{m}-th order given in a coordinate system with origin at the center
of cylinder $\alpha$. Due to the boundary conditions at infinity, the scattered field is expressed
in terms of outgoing cylindrical Hankel functions of second kind, $H_{m}^{(2)}(x)$. In the MSM
formalism the coefficients $b_{m}^{(\alpha)}$ (and implicitly the total field at the FF) can be
calculated by solving a system of linear equations,
$\mathbf{S}_{\omega}\mathbf{b}_{\omega}=\mathbf{a}_{\omega}$ (the vectors $\mathbf{a}_{\omega}$
and $\mathbf{b}_{\omega}$ depend only on the coefficients $a_{m}^{(0)}$ and $b_{m}^{(\alpha)}$,
respectively), with the scattering matrix, $\mathbf{S}_{\omega}$, being fully determined by the
distribution and material parameters of the nanowires \cite{ftm94josaa}.

In the second step of the nonlinear MSM method one determines the field at the SH, generated by
the nonlinear polarization,
$\mathbf{P}_{\mathrm{nl}}(\mathbf{r};\Omega)=\epsilon_{0}\mathbf{\hat{\chi}}_{s}^{(2)}:\mathbf{E}(\mathbf{r};\omega)\mathbf{E}(\mathbf{r};\omega)
\delta(\mathbf{r}-\mathbf{r}_{s})+\alpha [\mathbf{E}(\mathbf{r};\omega) \cdot
\nabla]\mathbf{E}(\mathbf{r};\omega) + \beta \mathbf{E}(\mathbf{r};\omega) [\nabla \cdot
\mathbf{E}(\mathbf{r};\omega)] + \gamma \nabla [\mathbf{E}(\mathbf{r};\omega) \cdot
\mathbf{E}(\mathbf{r};\omega)]$, where $\Omega=2\omega$. Here, the first term is a local dipole
surface polarization and the last three terms describe the nonlocal, bulk polarization
\cite{h91book}. If the free electrons in the metal are described by the Drude model then
$\alpha=0$, $\beta = \epsilon_0 e/(2 m_0 \omega^2)$, and $\gamma = \beta[1 -
\epsilon_r(\omega)]/4$, where $e$ and $m_{0}$ are the electron charge and mass, respectively, and
$\epsilon_r(\omega)=1 - \omega_p^2/[\omega (\omega + i \nu)]$ is the Drude permittivity. For Au
plasma frequency, $\omega_{p}=1.37\times10^{16}~\mathrm{s^{-1}}$, and damping frequency,
$\nu=4.05\times10^{13}~\mathrm{s^{-1}}$ \cite{oba85ao}. The surface susceptibility
$\mathbf{\hat{\chi}}_{s}^{(2)}$ of homogeneous, isotropic centrosymmetric media, such as noble
metals, has three independent components, which for Au are
$\mathbf{\hat{\chi}}_{s,\perp\perp\perp}^{(2)}=1.59\times10^{-18}~\mathrm{m^{2}/V}$,
$\mathbf{\hat{\chi}}_{s,\parallel\parallel\perp}^{(2)} =
\mathbf{\hat{\chi}}_{s,\parallel\perp\parallel}^{(2)}=4.63\times10^{-20}~\mathrm{m^{2}/V}$, and
$\mathbf{\hat{\chi}}_{s,\perp\parallel\parallel}^{(2)} = 0$ \cite{ktr04jap}.

The total field at the SH, $\vert \Psi_{\mathrm{tot}}\rangle$, is the sum between the scattered
field, $\vert \Psi_{\mathrm{sc}}\rangle$, and the source field, $\vert
\Psi_{\mathrm{src}}\rangle$, generated by the nonlinear polarization
$\mathbf{P}_{\mathrm{nl}}(\mathbf{r};\Omega)$. The source field obeys the inhomogeneous Helmholtz
equation and thus it can be calculated as the convolution between the Green function of the 2D
Helmholtz equation, $G_{2D}(r)=-(i/4)H_{0}^{(2)}(kr)$, and the corresponding source term:
$\langle\mathbf{r}\vert \Psi_{\mathrm{src}}\rangle=-i\Omega
G_{2D}\otimes\left[\left(\nabla\times\mathbf{P}_{\mathrm{nl}}\right)\cdot \mathbf{e}_{z}\right]$.
Similarly to Eq. (\ref{eq:scattered}), the scattered field can be expanded as
\begin{equation}
  \label{eq:scattered_nl}
  \vert \Psi_{\mathrm{sc}}\rangle = \sum_{\alpha=1}^{N}\sum_{m=-\infty}^{\infty} B_{m}^{(\alpha)} \vert
  Hm\rangle_{\alpha},
\end{equation}
where the coefficients $B_{m}^{(\alpha)}$ can be determined by solving a linear system of
equations, $\mathbf{S}_{\Omega}\mathbf{B}_{\Omega}=\mathbf{A}_{\Omega}$. Here, the vector
$\mathbf{A}_{\Omega}$ is known, being determined by the expansion coefficients of the source
field, and $\mathbf{S}_{\Omega}=\mathbf{S}_{\Omega=2\omega}$ is the nonlinear scattering matrix of
the system (a detailed description of this method can be found in Ref. \cite{bp10prb}).

\textit{Properties of nonlinear plasmonic WGMs}.---We have used this approach to explore the
physical properties of nonlinear plasmonic WGMs and how they are influenced by the cavity geometry
and the electromagnetic properties of the surrounding medium. To begin with, we show in Figs.
\ref{fig:geometry}(b) and \ref{fig:geometry}(c) a generic example of a nonlinear WGM of a
hexagonal plasmonic cavity. This mode has been computed by using a multipole source $\vert
\psi_{\mathrm{ex}}\rangle = \vert J2\rangle$, namely, all coefficients in Eq. (\ref{eq:incoming})
were set to zero, except the one with $m= m_0=2$. Note that the field profiles of the WGM in Fig.
\ref{fig:geometry} are the scattered fields at the FF and SH, which do not depend on the
particular way in which the mode is excited. However, by properly setting the value of $m_{0}$ one
can excite WGMs with specific symmetry, \textit{i.e.} modes that are invariant to symmetry
operations of the $\mathcal{C}_{m_{0}}$ symmetry group of $m_{0}$-fold rotations.
\begin{figure}
\includegraphics[width=7.5cm]{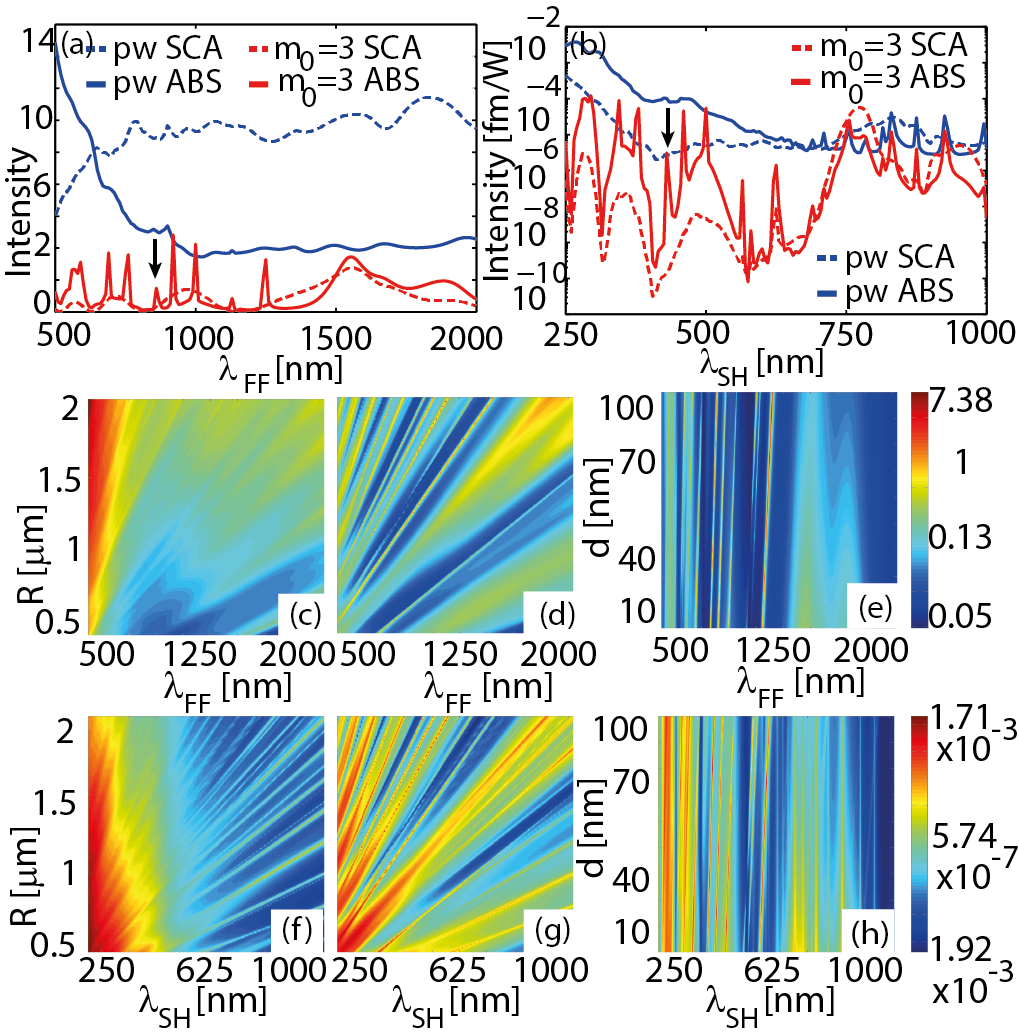}
\caption{\label{fig:spectra} Panels (a) and (b) show the absorption and scattering cross sections for a 6 cylinder cavity with $R=1500~\mathrm{nm}$, excited by a plane wave or a multipole source with$m_{0}=3$. Arrows correspond to the WGM in Fig. \ref{fig:fields}. (c)-(h) Dispersion maps of the absorption cross section \textit{vs}. $R$ and $d$: (c), (d) and (e), (h) correspond to a plane wave excitation and a multipole source with $m_0=3$, respectively. Panels on the left (right) correspond to the FF (SH).}
\end{figure}

The most efficient way to find WGMs of the cavity is to determine the resonances in the absorption
and scattering cross-section spectra, at both the FF and SH. Figures \ref{fig:spectra}(a) and
\ref{fig:spectra}(b) show the cross-section spectra corresponding to a hexagonal cavity that is
excited by either a plane wave or a multipole source with $m_{0}=3$. As the strong field
confinement leads to increased optical absorption in the metal, absorption spectra reveal more
information about the modes of the cavity as compared to the scattering spectra. Indeed, it can be
seen in Figs. \ref{fig:spectra}(a) and \ref{fig:spectra}(b) that the absorption spectra contain a
larger number of spectral peaks. Another important idea revealed by these spectra is that
multipole sources excite many more modes than plane waves. Because plane waves cannot excite WGMs
(they do not carry angular momentum) we expect that the additional resonances seen in the spectra
corresponding to multipole source excitation are associated to WGMs. Equally important is the fact
that in most cases a spectral resonance at the FF has a counterpart in the SH spectra, which means
that they correspond to two-component (multi-color) nonlinear modes. This is primarily because a
strong field at the FF generates a large nonlinear polarization and consequently an enhanced field
and optical absorption at the SH.

\begin{figure}
\includegraphics[width=7.5cm]{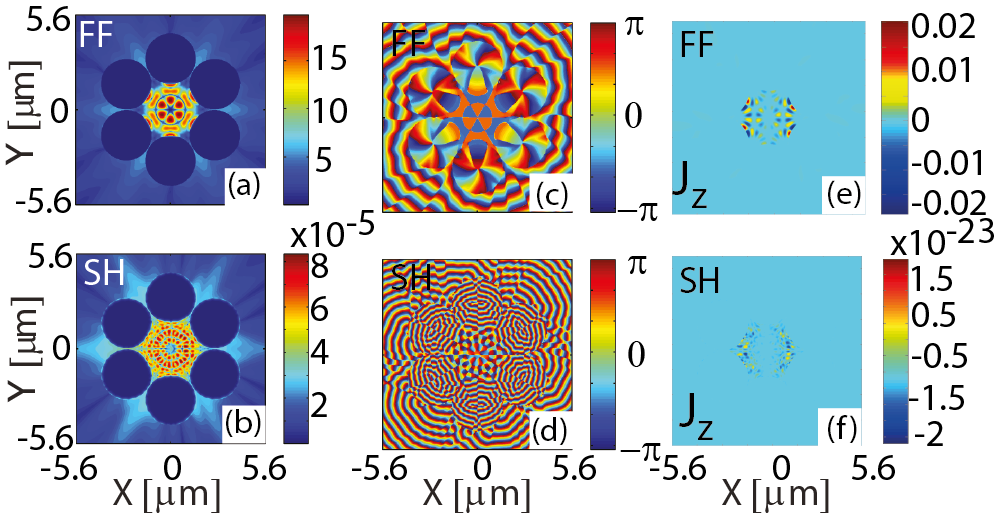}
\caption{\label{fig:fields} Spatial distribution of $\vert\mathbf{E}\vert$ (panels (a) and (b)), phase of $H_{z}$ (panels (c) and (d)), and $\mathcal{J}_{z}$ (panels (e) and (f)) for a hexagonal cavity with $R=1500~\mathrm{nm}$ and $d=10~\mathrm{nm}$. The resonance wavelength is $\lambda_{\mathrm{FF}}=863~\mathrm{nm}$ and $m_{0}=3$.}
\end{figure}
Further insights into the nature of the cavity modes are provided by the dispersion maps of the
absorption spectra shown in Figs. \ref{fig:spectra}(c)-\ref{fig:spectra}(h). Thus, is can be seen
that while the resonance wavelength of all cavity modes varies with the radius of the nanowires,
it depends on the separation distance only for certain modes. The modes which are independent on
$d$ represent multipole plasmon modes of single nanowires whereas the modes whose resonance
wavelength varies with $d$ are plasmonic cavity modes. Of these latter ones, plasmonic WGMs are
those that do not couple with plane waves, \textit{i.e.}, they are only excited by multipole
sources. This shows that one can find WGMs with specific symmetry properties by simply setting the
order of the multipole source, $m_{0}$.

\begin{figure}[b]
\includegraphics[width=7.5cm]{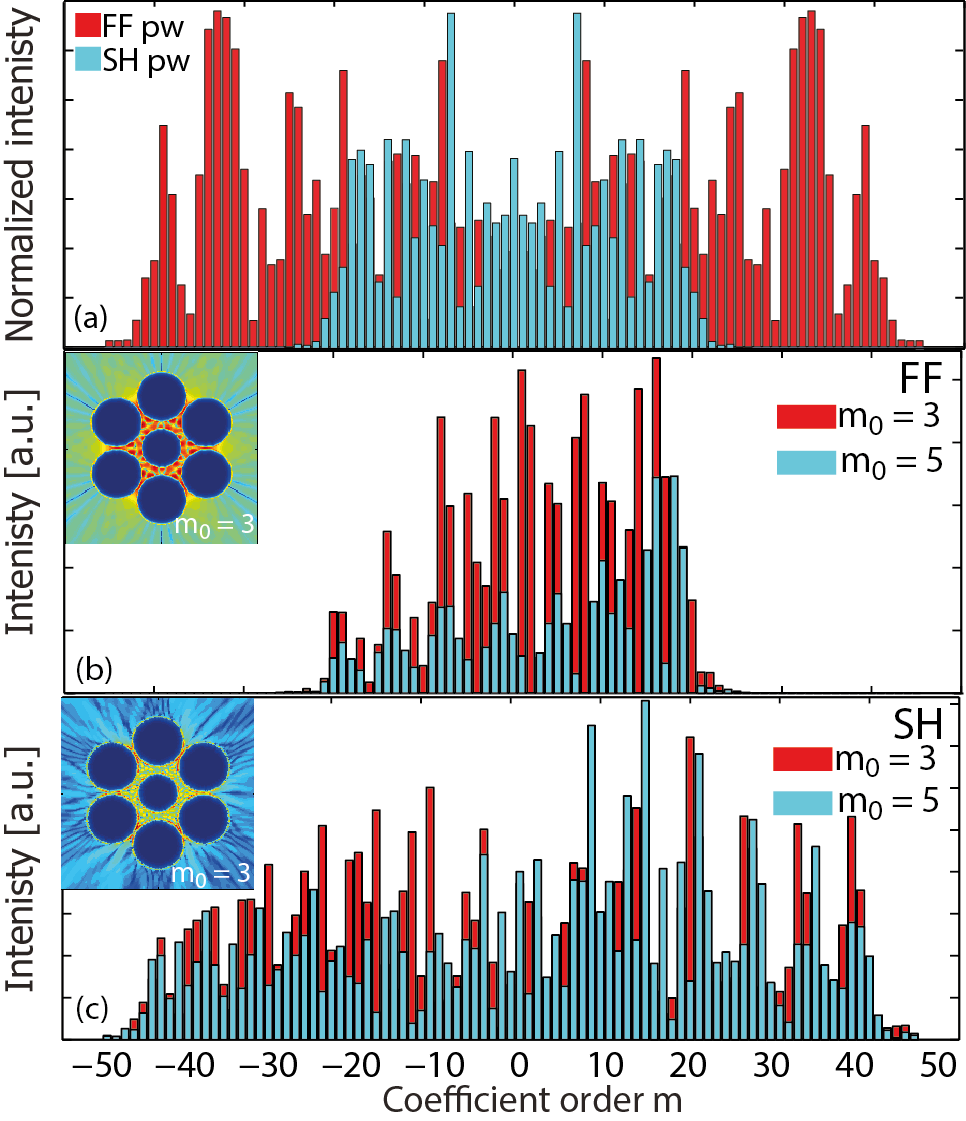}
\caption{\label{fig:fourier} Distribution of scattered field Fourier coefficients for a six cylinder cavity with $R=1500~\mathrm{nm}$, $R_i=1200~\mathrm{nm}$, and $d=10~\mathrm{nm}$ at $\lambda_{\mathrm{FF}}=815~\mathrm{nm}$ corresponding (a) to a plane wave excitation and (b), (c) multipole sources with $m_0=3$ and $m_0=5$ at the FF and SH, respectively.}
\end{figure}
The symmetry properties of the nonlinear WGMs are illustrated by the spatial distribution of the
optical field and its phase, as shown in Figs. \ref{fig:fields}(a)-\ref{fig:fields}(d). Thus, it
can be seen that the field at the SH has twice as many nodes on circles centered at the origin as
compared to the field at the FF, a property that is not specific to this particular nonlinear WGM.
In particular, the phase of the magnetic field indicates the existence of an optical vortex
located at the center of the cavity, its topological charge being $\sigma=m_{0}=3$ at the FF and
$\sigma=2m_{0}=6$ at the SH.

In order to quantify the angular momentum carried by the WGMs we introduce an effective azimuthal
modal number, $M_{\mathrm{eff}}=\omega \mathcal{J}_{z}/\mathcal{U}$, where $\mathcal{J}_{z}$ and
$\mathcal{U}$ are the optical cycle-averaged $z$-component of the total angular momentum and
electromagnetic energy, per unit length, respectively. For the TE polarization they are given by:
\begin{subequations}\label{fieldquant}
\begin{eqnarray}
  \label{J}\mathcal{J}_{z} &=& -\frac{1}{2c^{2}}\int \textsf{Re}\left(rE_{r}H_{z}^{*}\right)d\mathbf{r}, \\
  \label{U}\mathcal{U} &=& \frac{1}{4}\int \textsf{Re}\left[\epsilon(\mathbf{r})\left(\vert E_{r}\vert^{2}+\vert E_{\phi}\vert^{2}\right)+\mu_{0} \vert H_{z}\vert^{2}\right]d\mathbf{r}.
\end{eqnarray}
\end{subequations}
Our choice for the definition of $M_{\mathrm{eff}}$ is guided by the fact that
$M_{\mathrm{eff}}=m$ for a multipole field with axial index $m$ \cite{n08book}. Note, however,
that for such a field both $\mathcal{J}_{z}$ and $\mathcal{U}$ are infinite. From a physical point
of view, $M_{\mathrm{eff}}$ represents the average angular momentum carried by a photon in the
mode, in units of $\hbar$. Figures \ref{fig:fields}(e) and \ref{fig:fields}(f) show that the total
$\mathcal{J}_{z}\neq0$ at both the FF and SH, the calculated effective azimuthal modal numbers
being $M_{\mathrm{eff}}^{\omega}=0.0514$ (FF) and $M_{\mathrm{eff}}^{\Omega}=0.1120$ (SH),
\textit{i.e.} $M_{\mathrm{eff}}^{\Omega}=2M_{\mathrm{eff}}^{\omega}$ within numerical errors.
These results show that, remarkably, plasmonic systems can have multi-color nonlinear plasmonic
modes caring fractional angular momentum at both frequencies, meaning that such peculiar modes are
not restricted to the linear regime \cite{rfm12prb}.

\begin{figure}[t]
\includegraphics[width=8cm]{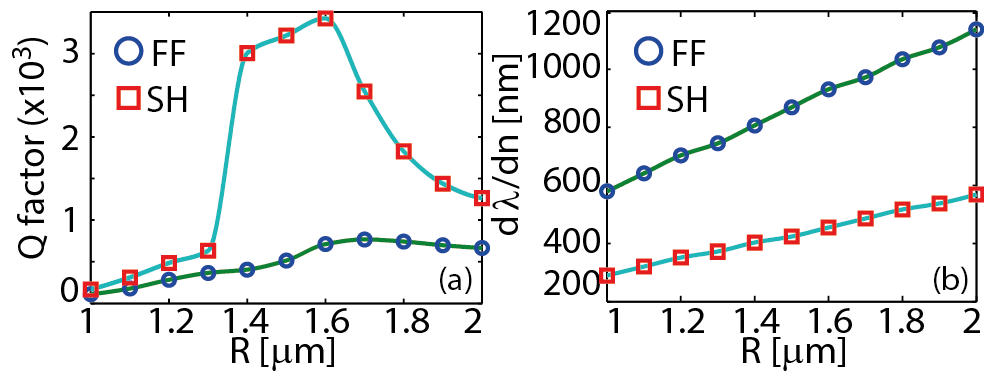}
\caption{\label{fig:applications} (a), (b) Dependence of (a) $Q$-factor and (b) sensitivity $S$ on $R$. The WGM and cavity parameters are as in Fig. \ref{fig:fields}.}
\end{figure}
More physical insights into the properties of the WGMs, \textit{i.e.} the origin of the fractional
character of their angular momentum, can be inferred by analyzing the angular dependence of the
optical field, which is determined by the spectrum of the Bessel-Fourier coefficients of the field
scattered at the FF and SH. Thus, one can easily demonstrate that in the case of an incident plane
wave with the wave vector along one of the symmetry axes of the cavity the scattering coefficients
expressed in the coordinate system with the origin at the center of the cavity satisfy the
symmetry relations $b_{-m}=(-1)^{m}b_{m}^{*}$ and $B_{-m}=(-1)^{m}B_{m}^{*}$ [see Fig.
\ref{fig:fourier}(a)]. Hence, the cylindrical multipoles with axial number $m$ and $-m$ contribute
equally to the scattered field, at both the FF and SH. Therefore, in this case
$\mathcal{J}_{z}=0$, meaning that no WGMs can be excited. These symmetry relations no longer hold
for a multipole source excitation ($m_{0}\neq 0$), implying that in this case plasmonic modes with
finite angular momentum can exist. This modal asymmetry is illustrated by the spectra of the
Bessel-Fourier coefficients shown in Figs. \ref{fig:fourier}(b) and \ref{fig:fourier}(c), which
correspond to $m_{0}=3$ and $m_{0}=5$, respectively. These spectra also show that, as expected, a
cavity with $p$-fold symmetry will predominantly scatter the field produced by a multipole source
with axial number $m_{0}$ into modes with $m=m_{0}\pm np$, with $n=0, 1, 2, \ldots$. In addition,
when $m_{0}=3$ the amplitudes of the scattering coefficients are larger than in the case of
$m_{0}=5$, which is explained by the fact that in the former case there is a stronger overlap
among the modes with the predominant contribution to the scattering process ($m=\pm3$ \textit{vs}.
$m=\pm1,\pm5$, respectively).

\textit{Applications of nonlinear WGMs to plasmonic sensors}.---Plasmonic cavities have optical
modes with extremely small modal volume and relatively large $Q$-factor, which makes them suitable
for ultra-compact sensing devices \cite{ia06stq,bp1011oenan}. To assess the potential of nonlinear
WGMs for sensing applications we calculated $Q$-factor and sensitivity, $S=d\lambda/dn$,
$n=\sqrt{\epsilon_{b}}$ being the background refractive index, of the WGM presented in Fig.
\ref{fig:fields}. The results of this analysis are shown in Fig. \ref{fig:applications}.

Because WGMs do not couple to the radiation continuum, one expects that they have large $Q$, a
conclusion validated by the results shown in Fig. \ref{fig:applications}(a). In fact, our
calculations suggest that nonlinear WGMs have larger $Q$ as compared to that of plasmonic cavity
modes with $m=0$ \cite{bp1011oenan}. Another unique feature of WGMs is that, unlike modes with
$m=0$, they can be used to probe at the nanoscale optical near-fields with specific chirality.
While $Q$-factor is relatively large at both the FF and SH, it is larger at the SH, reaching a
particularly large maximum value of $Q=3500$ for $R=1.6~\mathrm{\mu m}$. Equally important, Fig.
\ref{fig:applications}(b) shows that the sensitivity of the WGM is extremely large as well. In
particular, $S=1200~\mathrm{nm/RIU}$ for $R=2~\mathrm{\mu m}$, which is about three times larger
than recently reported sensitivities of sensors employing localized plasmon modes \cite{lmw10nl}.
Not surprisingly, the sensitivity at the FF is about twice as large as that at the SH, which is
the same as the ratio of the two wavelengths.

To conclude, we have demonstrated that plasmonic microcavities made of metallic nanowires have
nonlinear WGMs with particularly large $Q$-factors and sensitivities. These unique properties, of
significant practical relevance, are a consequence of the fact that nonlinear WGMs do not couple
to the radiation continuum. Importantly, nonlinear plasmonic WGMs can have fractional azimuthal
modal index, which opens up the possibility to generate at the nanoscale optical fields with
fractional angular momentum. Applications of WGMs to nano-optics, including plasmonic sensors,
have also been discussed.

\textit{Aknowledgements}.---The authors acknowledge the use of the UCL \textit{Legion} High Performance Computing Facility
(Legion@UCL), and associated support services, in the completion of this work. This work was
supported by the EPSRC.

\end{document}